\documentclass [12pt] {article}
\usepackage{axodraw}

\catcode`@=12
\newcommand{\beq}{\begin{equation}}
\newcommand{\eeq}{\end{equation}}

\newcommand{\bea}{\begin{eqnarray}}
\newcommand{\eea}{\end{eqnarray}}

\begin{document}
\thispagestyle{empty}
\begin{flushright} UCRHEP-T390\\
May 2005\
\end{flushright}
\vspace{0.5in}
\begin{center}
{\LARGE \bf Aspects of the Tetrahedral\\ Neutrino Mass Matrix\\}
\vspace{1.5in}
{\bf Ernest Ma\\}
\vspace{0.2in}
{\sl Physics Department, University of California, Riverside,
California 92521, USA\\}
\vspace{2in}
\end{center}

\begin{abstract}
The four-parameter tetrahedral neutrino mass matrix introduced earlier by the 
author is studied in two specific limits, both having only two parameters and 
resulting in $\theta_{13}=0$, $\theta_{23}=\pi/4$, and $\tan^2 \theta_{12} = 
1/2$.  One limit corresponds to a recent proposal which predicts a normal 
ordering of neutrino masses; the other is new and allows both inverted 
and normal ordering.
\end{abstract}

\newpage
\baselineskip 24pt

The non-Abelian discrete symmetry group of the even permutation of 4 objects 
has 12 elements and is called $A_4$.  It is also the symmetry group of the 
tetrahedron and has been shown to be useful for understanding the family 
structure of charged-lepton and neutrino mass matrices \cite{mr01,bmv03}. 
A modified version \cite{m04} of this model was proposed a year ago by the 
author and has 4 parameters which allow arbitrary neutrino masses while 
predicting both $\theta_{23}$ and $\theta_{12}$ as functions of $\theta_{13}$. 
For $0 < |U_{e3}| < 0.16$, this implies $1 > \sin^2 2 \theta_{23} > 0.94$ and 
$0.5 < \tan^2 \theta_{12} < 0.52$ respectively.  At that time, the central 
value of $\tan^2 \theta_{12}$ in a global fit of all neutrino data 
\cite{gg04} was near 0.39, but now (with the most recent SNO analysis 
\cite{sno05}) it is given rather by \cite{sv05}
\begin{equation}
\tan^2 \theta_{12} = 0.45 \pm 0.05.
\end{equation}
This means that the prediction of Ref.~\cite{m04} is in much better shape 
and deserves a second look.

In the basis where the charged-lepton mass matrix is diagonal, it was shown 
in Ref.~\cite{m04} that a particular application of $A_4$ results in
\begin{equation}
{\cal M}_\nu = \pmatrix{a+(2d/3) & b-(d/3) & c-(d/3) \cr b-(d/3) & c+(2d/3) & 
a-(d/3) \cr c-(d/3) & a-(d/3) & b+(2d/3)}.
\end{equation}
If $b=c$, then this has the solution
\begin{equation}
\pmatrix{\nu_1 \cr \nu_2 \cr \nu_3} = \pmatrix{\sqrt{2/3} & -\sqrt{1/6} & 
-\sqrt{1/6} \cr \sqrt{1/3} & \sqrt{1/3} & \sqrt{1/3} \cr 0 & -\sqrt{1/2} & 
\sqrt{1/2}} \pmatrix{\nu_e \cr \nu_\mu \cr \nu_\tau},
\end{equation}
with
\begin{equation}
m_1 = a-b+d, ~~~ m_2 = a+2b, ~~~ m_3 = -a+b+d.
\end{equation}
Hence $\theta_{13}=0$, $\theta_{23}=\pi/4$, and $\tan^2 \theta_{12}=1/2$. 
This pattern is reminiscent of the $\pi-\eta-\eta'$ system in hadronic 
physics and was conjectured \cite{hps02} to be applicable in the neutrino 
sector as well.  It is consistent with all present neutrino-oscillation 
data.

Consider now the 3 neutrino mass eigenvalues of Eq.~(4).  With the 3 
parameters $(a,b,d)$, it is clear that $m_{1,2,3}$ may be chosen to fit 
whatever experimental values of $|m_2|^2-|m_1|^2$ and $|m_3|^2-|m_2|^2$ 
are necessary.  In other words, this model has no prediction on neutrino 
masses.  However, there are 2 interesting special cases of Eq.~(4) which 
have definite predictions: (I) $b=0$, and (II) $a=0$.

Case (I) has
\begin{equation}
m_1 = a+d, ~~~ m_2 = a, ~~~ m_3 = -a+d,
\end{equation}
and was obtained recently in a detailed model \cite{af05} which solves the 
vacuum alignment problem inherent in the Higgs sector used to obtain 
Eq.~(2) by appealing to extra dimensions.  It also avoids the problem of 
setting $b=c$ which is impossible to maintain by a symmetry if either 
parameter is nonzero.  Since $\Delta m^2_{sol}$ is much smaller than 
$\Delta m^2_{atm}$ experimentally, Case (I) implies that
\begin{equation}
|d| \simeq -2|a|\cos \phi,
\end{equation}
where $\phi$ is the relative phase between $a$ and $d$.  Hence
\begin{equation}
|m_{1,2}|^2 \simeq |a|^2, ~~~ |m_3|^2 \simeq |a|^2 (1 + 8 \cos^2 \phi), 
\end{equation}
requiring thus a normal ordering of neutrino masses. The $\nu_e$ kinematical 
mass is then given by
\begin{equation}
|m_{\nu_e}|^2 \simeq |m_{1,2}|^2 \simeq {\Delta m^2_{atm} \over 8 \cos^2 \phi},
\end{equation}
and the effective $\nu_e$ mass in neutrinoless double beta decay is
\begin{equation}
|m_{ee}| = |a+(2d/3)| = {1 \over 3} |\Delta m^2_{atm}|^{1/2} 
[(9/8\cos^2 \phi)-1]^{1/2},
\end{equation}
resulting in the interesting relationship
\begin{equation}
|m_{\nu_e}|^2 \simeq |m_{ee}|^2 + \Delta m^2_{atm}/9.
\end{equation}

Case (II) has
\begin{equation}
m_1 = -b+d, ~~~ m_2 = 2b, ~~~ m_3 = b+d,
\end{equation}
and has not been considered before.  It requires
\begin{equation}
|d| \simeq |b| (\cos \phi + \sqrt{3+\cos^2 \phi}),
\end{equation}
where $\phi$ is the relative phase between $b$ and $d$.  Hence
\begin{equation}
|m_{1,2}|^2 \simeq 4|b|^2, ~~~ |m_3|^2 \simeq 4|b|^2 + 4|b||d|\cos \phi 
\simeq 4|b|^2 [1+\cos \phi(\cos \phi + \sqrt{3+\cos^2 \phi})].
\end{equation}
For $\cos \phi = 1$, $|m_3|^2 - |m_{1,2}|^2 \simeq 12|b|^2$ (normal 
ordering) and $|m_{ee}| = 2|b|$.  For $\cos \phi = -1$, $|m_3|^2 - 
|m_{1,2}|^2 \simeq -4|b|^2$ (inverted ordering) and $|m_{ee}| = (2/3)|b|$.
In general,
\begin{equation}
|m_{ee}| = {1 \over 3} |\Delta m^2_{atm}|^{1/2} [(1+3/\cos^2 \phi)^{1/2} 
\pm 1]^{1/2},
\end{equation}
where $\pm 1$ refers to $\cos \phi > 0$ or $< 0$, and
\begin{equation}
|m_{\nu_e}|^2 \simeq |m_{1,2}|^2 \simeq {\Delta m^2_{atm} \over \cos \phi
(\cos \phi + \sqrt {3+\cos^2 \phi})},
\end{equation}
resulting in the relationship
\begin{equation}
|m_{\nu_e}|^2 \simeq 3|m_{ee}|^2 - (2/3) \Delta m^2_{atm}.
\end{equation}
By choosing $\cos \phi$ near zero, it is clear that the present experimental 
upper bound of about 0.3 eV for $|m_{ee}|$ may be reached, in which case 
the 3 neutrino masses are nearly degenerate.  This is also possible in 
Case (I).

In Case (I), $b=c=0$ is a natural limit of the symmetry.  In Case (II), 
$a=0$ is a natural limit but $b = c \neq 0$ is not.  However for $b \neq c$, 
it was shown in Ref.~\cite{m04} that the experimental bound $|U_{e3}| < 0.16$ 
limits the deviation of $\tan^2 \theta_{12}$ from 0.5 to only 0.52.  In other 
words, if extended to allow $b \neq c$, Case (II) does not predict all three 
angles, but given one, it does predict the other two.

In conclusion, two interesting two-parameter descriptions of the neutrino mass 
matrix have been discussed, each with the mixing matrix of Eq.~(3).  One 
admits only a normal ordering of neutrino masses and predicts 
Eq.~(10); the other allows inverted as well as normal ordering and 
predicts Eq.~(16).

%\newpage

\begin{center} {\bf Acknowledgements}\\
\end{center}

I am grateful for the hospitality of J. W. F. Valle and others at the 
Instituto de 
Fisica Corpuscular, Valencia, Spain during a recent visit where this work 
was initiated.  My research was supported in part by the U.~S.~Department of 
Energy under Grant No.~DE-FG03-94ER40837.

\bibliographystyle{unsrt}

\end{document}